\documentclass[%
 reprint,
superscriptaddress,
 amsmath,amssymb,
 aps,
]{revtex4-2}

\usepackage{graphicx}
\usepackage{dcolumn}
\usepackage{bm}
\usepackage{float}

\begin{document}

\preprint{APS/123-QED}

\title{Reconstruction of ultrafast exciton dynamics with a phase-retrieval algorithm
}

\author{Bruno Moio}
    \thanks{These two authors contributed equally}
    \email{bruno.moio@polimi.it}
    \affiliation{Department of Physics, Politecnico di Milano, 20133 Milano, Italy}
    \affiliation{Institute for Photonics and Nanotechnologies, IFN-CNR, 20133 Milano, Italy}
\author{Gian Luca Dolso}
    \thanks{These two authors contributed equally}
    \affiliation{Department of Physics, Politecnico di Milano, 20133 Milano, Italy}
\author{Giacomo Inzani}%
    \affiliation{Department of Physics, Politecnico di Milano, 20133 Milano, Italy}
\author{Nicola Di Palo}%
    \affiliation{Department of Physics, Politecnico di Milano, 20133 Milano, Italy}
    \affiliation{Institute for Photonics and Nanotechnologies, IFN-CNR, 20133 Milano, Italy}
\author{Shunsuke A. Sato}%
    \affiliation{Center for Computational Sciences, University of Tsukuba, Tsukuba 305-8577, Japan}
    \affiliation{Max Planck Institute for the Structure and Dynamics of Matter, 22761 Hamburg, Germany}%
\author{Rocío Borrego-Varillas}%
    \affiliation{Institute for Photonics and Nanotechnologies, IFN-CNR, 20133 Milano, Italy}
\author{Mauro Nisoli}%
    \affiliation{Department of Physics, Politecnico di Milano, 20133 Milano, Italy}
    \affiliation{Institute for Photonics and Nanotechnologies, IFN-CNR, 20133 Milano, Italy}
\author{Matteo Lucchini}%
    \email{matteo.lucchini@polimi.it}
    \affiliation{Department of Physics, Politecnico di Milano, 20133 Milano, Italy}
    \affiliation{Institute for Photonics and Nanotechnologies, IFN-CNR, 20133 Milano, Italy}

\date{\today}

\begin{abstract} 
The first step to gain optical control over the ultrafast processes initiated by light in solids is a correct identification of the physical mechanisms at play. Among them, exciton formation has been identified as a crucial phenomenon which deeply affects the electro-optical properties of most semiconductors and insulators of technological interest. While recent experiments based on attosecond spectroscopy techniques have demonstrated the possibility to observe the early-stage exciton dynamics, the description of the underlying exciton properties remains non-trivial. In this work we propose a new method called extended Ptychographic Iterative engine for eXcitons (ePIX), capable of reconstructing the main physical properties which determine the evolution of the quasi-particle with no prior knowledge of the exact relaxation dynamics or the pump temporal characteristics. By demonstrating its accuracy even when the exciton dynamics is comparable to the pump pulse duration, ePIX is established as a powerful approach to widen our knowledge of solid-state physics.

\end{abstract}

\maketitle
\onecolumngrid
\section{\label{sec:intro}Introduction}

The possibility to observe and control the ultrafast physical processes triggered by light in matter is a long-pursued goal of solid-state physics which will enable future device engineering along with the acquisition of superior performances \cite{Kruchinin2018}. In this respect, the identification and understanding of the electron dynamics dictating the sudden energy redistribution after optical excitation are a prerequisite to achieve control. In the last decade, Attosecond Transient Absorption/Reflection Spectroscopy (respectively ATAS and ATRS) have been established as one of the most powerful tools to investigate the ultrafast optical response of metals \cite{volkov2019attosecond}, semiconductors \cite{schlaepfer2018attosecond,schultze2014attosecond,zurch2017direct,Zurch2017b,kaplan2018femtosecond,Attar2020,Buades2021} and insulators \cite{lucchini2016attosecond,mashiko2016petahertz,schultze2013controlling,mashiko2018multi,lucchini2021unravelling}. These spectroscopic techniques take advantage of the element-specificity of extreme-ultraviolet (XUV) photons to probe the ultrafast dynamics induced by the interaction of the solid sample with a few-cycle infrared (IR) pump pulse \cite{geneaux2019transient}. Particular attention has been recently drawn by the study of the dynamical response of excitons, as these quasi-particles find applications in many relevant technological areas, including optoelectronics, photonics and excitonics \cite{Koch2006,Scholes2006,Butov2017}. Due to the relatively high photon energy of the attosecond radiation, the first pioneering experiments focused on core excitonic states, i.e. quasi-particles formed by the Coulomb interaction between an electron in the conduction band and a core hole \cite{moulet2017soft,geneaux2020attosecond,lucchini2021unravelling}.\\
Typically, a few-femtosecond IR pulse drives the system out of equilibrium, while the optical response is probed by the attosecond XUV radiation. Information about the exciton dynamics is encoded in a differential spectrogram, obtained by measuring the spectrum of the transmitted/reflected XUV radiation as a function of the relative delay with the IR pump pulse. Unfortunately, it is not possible to directly access the exciton characteristics (\textit{e.g.} its lifetime, polarizability, etc.) from the measured spectrogram. A possible approach to overcome this limitation is to fit the experimental data using a mathematical model to estimate the real-time behaviour of the core-exciton state \cite{geneaux2020attosecond}. To describe comprehensively the time-dependent excitonic dipole several fitting parameters must be considered (\textit{i.e.}, phonon coupling, Auger decay, complex polarizability, and pump laser characteristics). Therefore, without a well-educated guess or an \textit{a-priori} knowledge of some of the above mentioned physical parameters the fitting procedure may be ill-posed, hindering a reliable convergence.

In this work, we propose a novel approach named extended Ptychographic Iterative engine for eXcitons (ePIX), to retrieve the ultrafast dynamics of a core-exciton state from a transient reflection spectrogram. This method requires no prior information about the dynamics of the excited state and no complete characterization of the driving IR field. Moreover, since the algorithm is based on a two-level atomic model, ePIX can be used to investigate the ultrafast physical processes happening in several other systems which can be framed within this representation, like atoms and molecules \cite{Herrmann2013,Drescher2019,Drescher2020}. Therefore, the proposed methodology is general and its applicability extends beyond the solid-state physics.\\
Our work is based on a reformulation of the Ptychographic Iterative Engine (PIE) algorithm and its extended version (ePIE) \cite{spangenberg2015time,spangenberg2015ptychographic,lucchini2015ptychographic} used to characterize the light pulses involved in attosecond streaking experiments \cite{itatani2002attosecond}. Compared to a 2D fitting procedure, we found that ePIX assures robust convergence also when no \textit{a-priori} information on the system physical parameters is available. In addition, the ePIX retrieval procedure does not assume any functional form of the exciton dynamics and can therefore retrieve an exciton time-dependent dipole with an arbitrary shape. While this is a big advantage when the time-dependent dipole constitutes the primary object of the study, a mathematical model is still needed if one aims to extract the exciton physical parameters (e.g. Auger decay) from the dipole time evolution, thus requiring a subsequent 1D fitting procedure. However, as we will discuss in the following, applying ePIX reconstruction and a subsequent 1D fit gives a more accurate reconstruction than a global 2D fitting procedure, especially when the observed dynamics unfolds on a time scale comparable to the pump pulse duration.\\

The work is organized as follows. In Section \ref{sec:theory} we present the atomic-like model that predicts the temporal evolution of the core-exciton ground state under the effect of an external electric field. Section \ref{sec:algorithm} describes the implementation of the reconstruction algorithm we developed. We present the results of the reconstruction applied to a show-case simulated ATRS trace in Section \ref{sec:results}, while the reconstruction output at each step of the algorithm are discussed in Section \ref{sec:discussion}.

\section{\label{sec:theory}Core-exciton transient reflectivity contribution}
\begin{figure*}[htbp]
    \centering\includegraphics[width=0.9\textwidth]{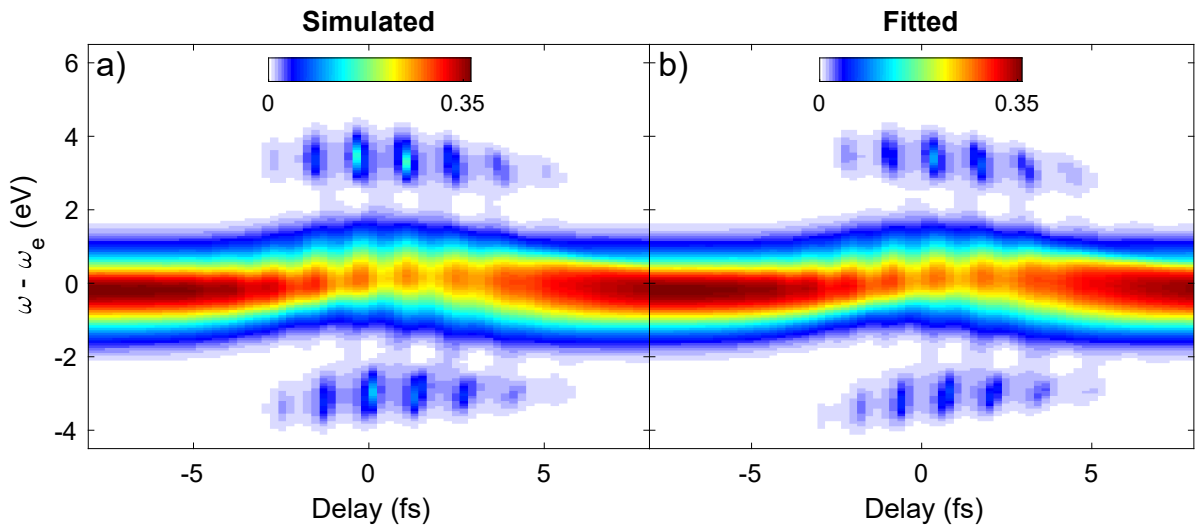}
    \caption{\label{fig:comp_nl}(a) Simulated  transient reflectance, $R(\omega, \tau)$, for an incidence angle of $\theta =$\,73.5$^{\circ}$, obtained by assuming an exciton Auger decay lifetime $T_e = 4$\,fs, a phonon coupling and frequency $M_0 = 0.3$\,PHz and $\omega_0 =0.15$\,rad\,PHz, respectively. The exciton frequency is $\omega_e = 84$\,rad\,PHz. The IR pulse has a peak intensity of 0.78\,V/\AA\, and couples with the system through a complex polarizability of ($2 - i2.5$)\,rad\,PHz\,\AA$^{2}$\,V$^{-2}$. The dipole amplitude is $k = 30$\,PHz. The wavelength of the IR field is set to 800\,nm, the full-width-half-maximum duration to 8\,fs, the carrier-envelope phase (CEP) to 0.4\,$\pi$ and its chirp rate to 0.05\,fs$^{-2}$. (b) Result of a 2D least-square non-linear fitting procedure. Despite the good qualitative agreement between the traces in (a) and (b), the physical parameters, listed in Tab.~\ref{tab:allresults}, are not accurately reconstructed.}
\end{figure*}
In an ATAS or ATRS experiment, the attosecond radiation creates the core exciton in the presence of the external IR driving field, which perturbs the system relaxation process. On the attosecond time scale the IR field can induce solid-like phenomena like the dynamical Franz-Keldysh effect \cite{Novelli2013}, which produces fast oscillations in the transient signal, related to the exciton nanometric motion \cite{lucchini2021unravelling}. The system response on the few-fs time scale is instead dominated by atomic-like effects, such as the optical Stark effect \cite{mysyrowicz1986dressed}, which transient signature encodes information on the exciton relaxation mechanism \cite{geneaux2020attosecond}. In this work we will concentrate on the latter class of phenomena as the study of the attosecond timing of the exciton dynamics requires a different experimental and theoretical approach which goes beyond the present scope. In particular, we will model the exciton with non-dispersive localized states, whose temporal evolution is described by an exponential decay due to an Auger relaxation process plus a Gaussian decay via phonon coupling \cite{mahan1977emission}, and concentrate only on the femtosecond dynamics. For the sake of simplicity, we analyse the evolution of the exciton in its ground state, neglecting other bright or dark excited states. As we will discuss in Section \ref{sec:conclusions}, this does not limit the validity and generality of our approach.\\

If we assume the exciton formation to be instantaneous and the XUV pulse to be considerably shorter than the pump pulse, so that it can be approximated with a delta function in time, the system electrical permittivity $\epsilon(\omega, \tau)$, as a function of the XUV photon frequency $\omega$ and of the delay $\tau$ between the XUV and IR pulses, can be expressed as follows \cite{moulet2017soft}:
\begin{equation}
\label{eq:model}
    \epsilon(\omega, \tau) = \epsilon_0+ k \int_{-\infty}^{+\infty} H(t)\, e^{-\frac{t}{T_e}} e^{-i\omega_e t} e^{\Phi(t)} e^{i\Psi_{IR}(t)} e^{i\omega t}\,dt
\end{equation}
where $\epsilon_0$ is the crystal electrical permittivity, $k$ is a constant accounting for the strength of the core-exciton response, $H(t)$ is the Heaviside function describing the instantaneous exciton formation, $T_e$ is the Auger lifetime of the excited state and $\hbar \omega_e$  is the excitation energy. The function $\Phi(t)$ governs the coupling with phonons and reads:
\begin{equation}
    \Phi(t)=-\frac{M_0^2}{\omega_0^2} \left[ \left( 2N+1\right) \left( 1 - \cos{\omega_0t}\right) -i\left( \omega_0t - \sin{\omega_0t}\right) \right].
\end{equation}
Here, $M_0$ is the phonon coupling coefficient, accounting for the strength of the phonon relaxation, $\hbar \omega_0$ is the phonon energy and $N$ is the phonon population, derived from the Boltzmann statistics. The term $\Psi_{IR}$, which depends on the IR field $E(t)$, describes the optical Stark effect as follows \cite{wirth2011synthesized,delone1999ac}:
\begin{equation}
\Psi_{IR} = -\left( \frac{\alpha}{2}-i\gamma \right) \int_0^t E^2 \left( t' - \tau \right)\,dt',
\end{equation}
where $\alpha/2$ and $\gamma$ represent the real and imaginary part of the polarizability of the core exciton.\\
Starting from $\epsilon(\omega, \tau)$, the system reflectance or absorbance can be evaluated through the Fresnel equations. Following the choice made in the most recent experiments \cite{geneaux2020attosecond,lucchini2021unravelling}, we concentrate on transient reflectance, $R(\omega, \tau)$. In case of $s$-polarized light, this quantity can be written as:
\begin{equation}\label{eq:fresnel}
    R(\omega, \tau) = \left| \frac{n_0 \cos{\theta} - \sqrt{\epsilon(\omega, \tau) - n_0^2 \sin^2{\theta}}}{n_0 \cos{\theta} + \sqrt{\epsilon(\omega, \tau) - n_0^2 \sin^2{\theta}}} \right|^2
\end{equation}
where $n_0$ is the refractive index of the surrounding medium, typically vacuum for experiments in the XUV spectral region (i.e. $n_0 =$\,1), and $\theta$ is the angle of incidence, measured with respect to the surface sample normal.\\
Figure~\ref{fig:comp_nl}(a) shows a transient reflectance calculated with this model for an ideal system similar to what reported in Ref. \cite{lucchini2021unravelling}: at an angle of incidence $\theta =73.5^{\circ}$ and assuming physically-meaningful values for the parameters involved \cite{moulet2017soft,almairac1974lattice} as listed in Tab.~\ref{tab:allresults}. For relatively large delays, the trace shows a peak around the excitonic transition, $\omega = \omega_e$, corresponding to the quasi-particle contribution to the total reflectivity. Around the pump-probe temporal overlap (about 0\,fs), we observe two major effects: (\textit{i}) a blueshift of the excitonic peak, clear signature of the optical Stark effect and a non-zero $\alpha$; (\textit{ii}) the formation of additional structures about 3\,eV above/below the excitonic peak, originating from the IR-field dressing of the excitonic resonance. Moreover, because of a finite $\gamma$, the reflectivity is modulated at twice the IR frequency.

\begin{table*}[htbp]
    \caption{\label{tab:allresults}%
    Simulated and retrieved values of the excitonic dipole $O(t)$ at the different stages during the ePIX algorithm. The indicated errors are the standard deviations arising from the results of four different reconstructions.
    }
    \begin{ruledtabular}
    \begin{tabular}{lccccc}
    \textrm{Parameter}&
    \textrm{Sim.}&
    \textrm{Fit}&
    \textrm{PIE I}&
    \textrm{ePIE}&
    \textrm{PIE II}\\
    \colrule
    $\alpha$ (rad\,PHz\,\AA$^{2}$\,V$^{-2}$)&
    $-$5&
    $-$4.6\,$\pm$\,0.013&
    $-$5.8\,$\pm$\,2.7&
    $-$5.3\,$\pm$\,0.75&
    $-$5.02\,$\pm$\,0.038\\
    $\gamma$ (rad\,PHz\,\AA$^{2}$\,V$^{-2}$)&
    2&
    1.5\,$\pm$\,0.0088&
    1.0\,$\pm$\,0.69&
    1.9\,$\pm$\,0.36&
    2.01\,$\pm$\,0.022\\
    $T_e$ (fs)&
    4&
    6\,$\pm$\,0.092&
    3.8\,$\pm$\,1&
    4.3\,$\pm$\,0.55&
    3.99\,$\pm$\,0.076\\
    $\omega_e$ (rad\,PHz)&
    84&
    84\,$\pm$\,0.00072&
    84.088\,$\pm$\,0.1&
    83.96\,$\pm$\,0.21&
    84.00\,$\pm$\,0.0083\\
    $M_0$ (PHz)&
    0.3&
    0.34\,$\pm$\,0.0053&
    0.25\,$\pm$\,0.058&
    0.32\,$\pm$\,0.054&
    0.30\,$\pm$\,0.0014\\
    $\omega_0$ rad\,PHz)&
    0.15&
    0.11\,$\pm$\,0.004&
    0.13\,$\pm$\,0.036&
    0.16\,$\pm$\,0.081&
    0.15\,$\pm$\,0.0057\\
    $k$ (PHz)&
    30&
    27.2\,$\pm$\,0.077&
    28.7\,$\pm$\,3.1&
    28.8\,$\pm$\,1.7&
    30.04\,$\pm$\,0.23\\
    \end{tabular}
    \end{ruledtabular}
\end{table*}

Calculating $R(\omega, \tau)$ from the real-time dipole is generally an easy task, but the reverse problem is not trivial and extracting the real-time evolution of the excited state from a transient reflectance measurement can be an hard task. A simple approach is to assume a parametric functional behaviour of the dipole moment and apply a non-linear fitting procedure to recover all the unknowns. This fitting approach is in general suitable for relatively simple dynamics; however, its application to retrieve the complex exciton behaviour may fail, without an \textit{a priori} knowledge of some of the physical parameters which describe the system, in particular when the investigated dynamics evolve on time scales close to the time duration of the pump pulse. Figure~\ref{fig:comp_nl}(b) shows the transient reflectance obtained by fitting the simulations of Fig.~\ref{fig:comp_nl}(a) with the same mathematical model employed for the direct calculations. Without knowing any of the system characteristics, the fitting free parameters are: the exciton parameters $T_{e}$, $\omega_{e}$, $M_{0}$, $\omega_{0}$, $k$, and the IR pulse duration, phase, and chirp. When such a high number of unknown parameters is employed, we found the 2D fitting procedure to be at times inaccurate. This is mainly because similar reflectivity traces can be generated within a range of physical parameters, causing the fitting procedure to converge to different solution depending on the initial guess. Indeed, while there is good qualitative agreement between the traces in Figs.~\ref{fig:comp_nl}(a)-(b), the retrieved parameters significantly differ from the ones used in the simulations (compare the second and third column in Table \ref{tab:allresults}). In particular, the errors in the reconstruction of the Auger lifetime and the phonon parameters seem to compensate, thereby hindering a reliable access to the information underlying the transient reflectivity trace. Therefore, using a simple fitting procedure to interpret attosecond transient reflectivity traces without additional information can be misleading. The new method we propose overcomes this limitation, allowing for the complete assessment of the core-exciton dynamics, regardless from its shape and complexity.

\section{\label{sec:algorithm}\MakeLowercase{e}PIX algorithm}

\begin{figure*}[htbp]
\centering\includegraphics[width=1\textwidth]{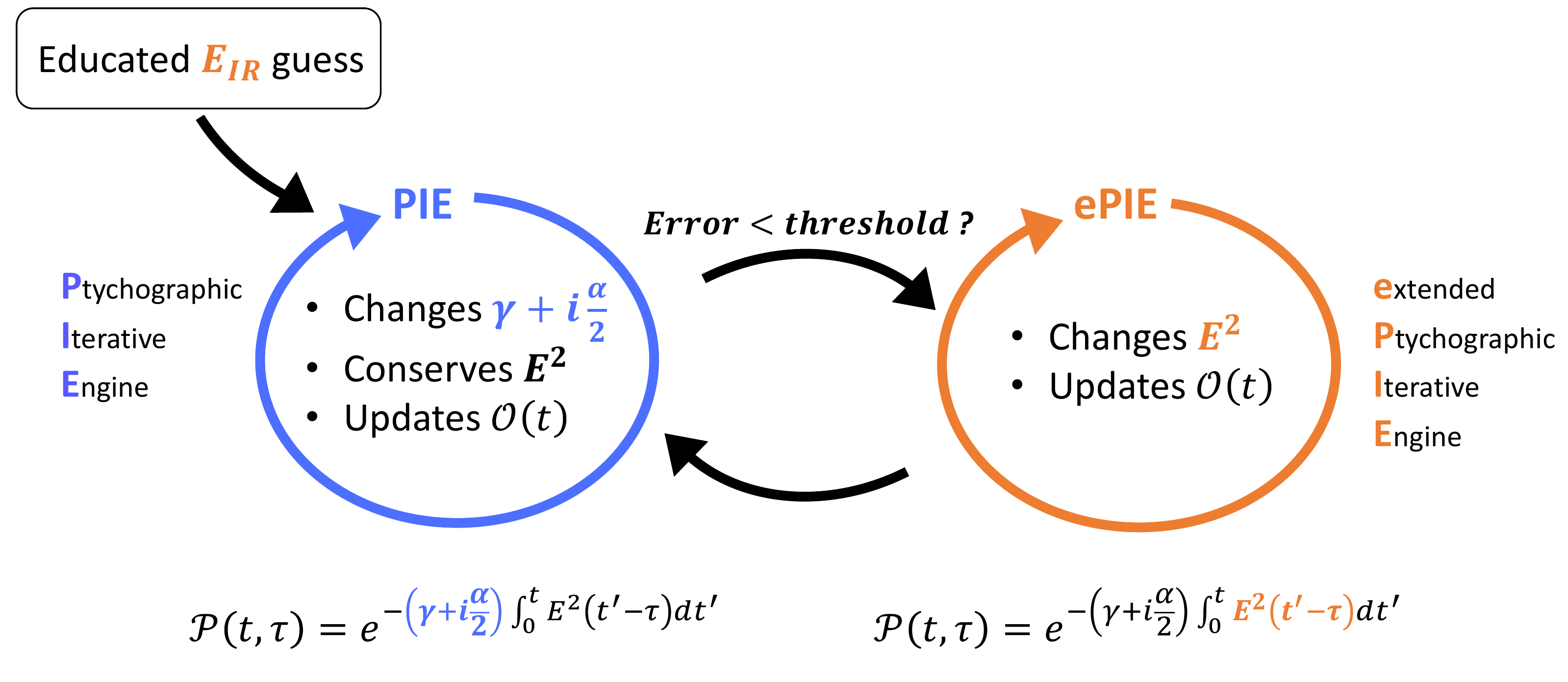}
    \caption{\label{fig:algorithm}Scheme of the ePIX reconstruction algorithm. The first stage (PIE) modifies the complex polarizability. After a certain reconstruction error has been achieved, the second stage (ePIE) is applied to correct also the IR field prediction. The result can then be fed back to the PIE stage as described in the main text, in case convergence requires further iterations. We found that a progression of three to five iterations is typically enough to reach convergence.}
\end{figure*}
In their standard implementation, ptychographic algorithms are based on the inversion of an internal product of the following form
\begin{align}\label{eq:spectrogram}
    S(\omega, \tau) &= \left| \int_{-\infty}^{+\infty} O(t)\,P(t, \tau)\,e^{i\omega t}\,dt\right|^{2}\\
    &= \left| \mathcal{F} \left\{ O(t) \cdot P(t,\tau)\right\}\right|^{2}
\end{align}
where $O(t)$ is the object to be reconstructed, $P(t)$ is called probe and $\mathcal{F}$ indicates the Fourier transform. When $P(t)$ is known, PIE \cite{McCallum1992} can be used to reconstruct the object $O(t)$. Instead, if both $P(t)$ and $O(t)$ are unknown, the ePIE algorithm is applied to retrieve both functions \cite{Thibault2009,Maiden2009,Maiden2017}. If we define the object to be the excitonic dipole
\begin{equation}\label{eq:dipole}
    O(t) = k \,H(t) \, e^{-\frac{t}{T_e}} e^{-i\omega_e t} e^{\Phi(t)}
\end{equation}
and the probe to equal the exponential describing the IR interaction
\begin{equation}\label{eq:probe}
    P(t,\tau) = e^{-\left( \gamma + i \frac{\alpha}{2}\right) \int_0^t E^2 \left( t' - \tau \right)\, dt'},
\end{equation}
then Eq.~(\ref{eq:model}) can be rewritten as
\begin{equation}
    \epsilon(\omega, \tau) = \epsilon_0 + \mathcal{F} \left\{ O(t) \cdot P(t,\tau)\right\},
\end{equation}
thus underlining the parallelism between Eq.~(\ref{eq:spectrogram}) and the excitonic contribution to the total permittivity. While this suggests the possibility to use ptychographic techniques to reconstruct the exciton dynamics, their applicability is not straightforward. Since $\epsilon (\omega, \tau)$ is subsequently related to the transient reflectivity by Eq.~(\ref{eq:fresnel}), in this case $O(t)$ and $P(t,\tau)$ are not related to the experimentally-measured quantity through a simple inner product as in Eq.~(\ref{eq:spectrogram}). In principle, one might think that this additional mathematical complexity could hinder convergence, but this is not the case. Since the electric dipole moment $O(t)$ respects the causality principle, leading to a Kramers-Krönig-consistent permittivity, the reconstruction problem is well-posed \cite{musfeldt1993method}. Nevertheless, to assure robust convergence, within the ePIX algorithm PIE and ePIE need to be applied in a non-conventional manner. The algorithm block diagram is shown in Fig.~\ref{fig:algorithm}.
At first, the IR pulse is assumed to be known and the PIE algorithm is run to retrieve the object and the exciton complex polarizability $\alpha/2 - i\gamma $. This assumption does not constitute a severe limitation as an educated guess for the IR pulse is typically available in attosecond experiments, \textit{e.g.} obtained as a side-product of the attosecond pulse characterization through attosecond streaking. In certain cases the IR pulse is measured simultaneously with the transient reflection/absorption experiment, assuring higher temporal resolution \cite{lucarelli2020novel}. As an alternative, the initial guess for the IR field can be derived from an independent optical measurement using, for example, the Frequency-Resolved Optical Gating (FROG)\cite{kane1993characterization} or the Spectral Phase Interferometry for Direct Electric-field Reconstruction (SPIDER)\cite{Iaconis1998} techniques. If the IR initial guess is accurate, this first PIE reconstruction is sufficient; if not, the excitonic polarizability is fixed to the current value and an ePIE reconstruction is performed to update both the object and the IR electric field. Following the iterative approach depicted in Fig.~\ref{fig:algorithm}, the result achieved by the ePIE stage can then be fed back to the PIE stage, if necessary.\\
To assure efficient convergence of the first reconstruction stage (PIE) we introduced two constraints on $P(t,\tau)$, which enforce the correct functional shape on the probe. This approach is analogous to the pure-phase constraint applied in the context of attosecond pulse ptychographic reconstruction \cite{lucchini2015ptychographic}. In particular, at each iteration we impose for the module ($|\cdot|$) and phase ($\angle\left\{\cdot\right\}$) of $P(t)$
\begin{equation}
    \begin{cases}
      \left| P(t) \right| - e^{-\gamma I} = 0\\
      2\angle\left\{P(t)\right\} + \alpha I = 0
    \end{cases}\label{constraint1}
\end{equation}
where
\begin{equation}
    I = \int_{0}^{t} E^2(t')\,dt'
\end{equation}
is the integral of the modulus squared of the IR field. These constraints are applied in the least-norm sense: at each iteration, the new values of $\alpha$ and $\gamma$ are evaluated by minimizing the norm of the left-hand side of Eqs.~(\ref{constraint1}). The updated value of the probe is thus given by:
\begin{equation}
    P(t,\tau) = e^{-\left( \gamma' + i \frac{\alpha'}{2}\right) \int_0^t E^2 \left( t' - \tau \right)\, dt'}.
\end{equation}
In this way it is possible to partially update the probe (polarizability) keeping the IR pulse unchanged. In this regard, our PIE stage differs from standard PIE applications, where the complete probe $P(t)$ does not change through the reconstruction.\\

Due to the complexity of the problem under examination, also the application of the second ePIE stage is not straightforward. The peculiar functional relation between reflectivity and permittivity, together with the presence of a complex polarizability which multiplies the electric field integral, can, in certain cases, prevent a fast convergence. To overcome this limitation, we follow an approach similar to that implemented by Keathley et al. \cite{Keathley2016} and assume the IR vector potential, $A(t)$, to have a known analytical form (i.e. Gaussian envelope and quadratic chirp). In this way, at each iteration the code needs to update only a reduced set of parameters (five in our case), which fully describe $A(t)$, increasing the convergence speed.\\

\section{\label{sec:results}Results}

In order to test the validity of ePIX, we applied our algorithm to the simulated trace of Fig.~\ref{fig:comp_nl}(a). Besides reproducing the behaviour of a physical problem, it represents an ideal test-case as all the physical parameters which can alter the transient optical response of the system are present. It is worth mentioning that the approach reported here is not limited to this specific physical case. We test ePIX against different optical field and exciton parameters and found that only few iterations of the PIE-ePIE stages (typically less than two) are required to converge with comparable results in all cases. We note that it is beneficial to end the ePIX procedure with a PIE stage, as this allows for a final refinement of the probe function, assuring better reconstruction than the 2D fitting procedure. The results we present are obtained by averaging a set of four different reconstructions of the same reflectivity trace, starting by different initial guesses of the involved quantities.

\begin{figure}[htbp]
\centering\includegraphics[width=0.46\textwidth]{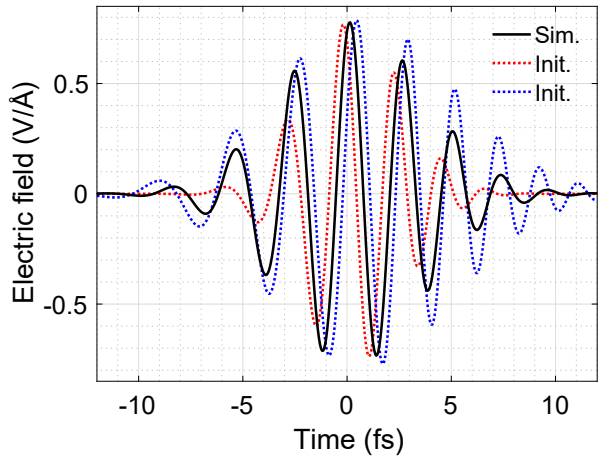}
    \caption{\label{fig:field}Simulated IR field (black curve) and two examples of IR initial guesses (dotted red and blue curves). For the simulated field, the wavelength is set to 800\,nm, the FWHM duration to 8\,fs, the carrier-envelope phase (CEP) to 1.26\,rad and its chirp rate to 0.05\,fs$^{-2}$. The initial guesses used in the reconstructions differ from the simulated field in duration, phase and chirp.}
\end{figure}

\subsection{\label{subsec:Istep}First stage: PIE}

\begin{figure*}[htbp]
\centering\includegraphics[width=0.9\textwidth]{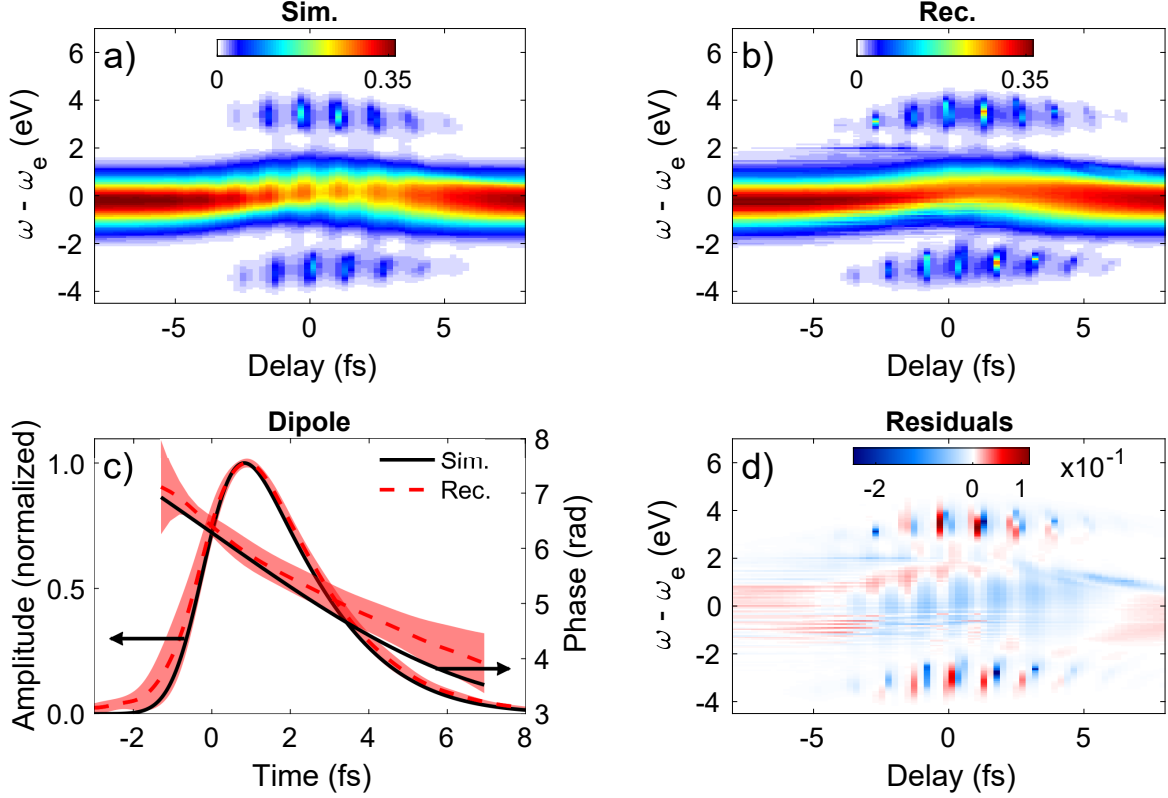}
    \caption{\label{fig:PIE1}Simulated, (a), and reconstructed, (b), transient reflectance trace, after the first PIE stage of the ePIX algorithm. (c) Input (solid black) and reconstructed (red dashed) excitonic dipole $O(t)$ amplitude and phase. (d) Difference between the reconstructed and simulated traces. The reconstructed dipole shows already a good degree of qualitative agreement after this first stage.}
\end{figure*}
The first step of the reconstruction consists in the application of the PIE algorithm described above. Assuming no \textit{a priori} knowledge on the exciton physical parameters, the initial guess of $O(t)$ is set to white noise while $\alpha$ and $\gamma$ are set to reasonable values, but different from the exact solutions, as summarized in Table~\ref{tab:init}. The input reflectivity trace is simulated with a Gaussian IR pulse having a FWHM time duration of 8\,fs, central wavelength of 800\,nm, a carrier-envelope phase (CEP) of 0.4\,$\pi$ and a chirp rate of 0.05\,fs$^{-2}$ (black curve in Fig.~\ref{fig:field}). Assuming a non-perfect knowledge of the IR, we perform four different reconstructions where the initial guesses are Gaussian pulses with a duration of either 5.6 or 10.3\,fs, a CEP of 2.0 or 0.5\,rad and a chirp rate of 0 or 0.1\,fs$^{-2}$ (two examples are reported in Fig.~\ref{fig:field}, dotted curves).
\begin{table}[htbp]
    \caption{\label{tab:init}%
    Initial guesses for the different parameters which characterize the probe and object. The initial guess for the dipole moment $O(t)$ is random noise. For the other parameters, two possible initial guesses are given: the four reconstructions explore different combinations of these initial guess values.
    }
    \begin{ruledtabular}
    \begin{tabular}{lcc}
    \textrm{Parameter}&
    \textrm{Simulated}&
    \textrm{Initial guess}\\
    \colrule
    $O(t)$                   & Eq.~(\ref{eq:dipole})                 & Random noise\\
    $\alpha$ (rad\,PHz\,\AA$^{2}$\,V$^{-2}$)                & $-$5              & $-$8 or $-$2\\
    $\gamma$ (rad\,PHz\,\AA$^{2}$\,V$^{-2}$)                & 2                 & 0.5\\
    FWHM (fs)                                               & 8                 & 5.6 or 10.3\\
    $\Phi$ (rad)                                            & 1.26              & 2 or 0.5\\
    CR (fs$^{-2}$)                                          & 0.05              & 0 or 0.1\\
    \end{tabular}
    \end{ruledtabular}
\end{table}
The scope of this approach is to test the accuracy of the reconstruction with a non-perfect knowledge of the IR pulse and to stress-test the reconstruction of the dipole moment parameters. 

Figure~\ref{fig:PIE1} shows the result of the first PIE stage, averaging over the four reconstructions performed. The reconstructed trace of Fig.~\ref{fig:PIE1}(b) already qualitatively agrees with the simulated input (Fig.~\ref{fig:PIE1}(a)) even if a non-perfect IR pulse is assumed (Fig.~\ref{fig:field}). The same holds for the reconstructed excitonic dipole (red dashed curve in Fig.~\ref{fig:PIE1}(c)). Within the reconstruction uncertainty (represented by the red shaded area which extends over two standard deviations), the dipole amplitude and phase largely agree with the simulation input (black solid curve). The overall agreement is further confirmed by the retrieved exciton physical parameters (see Section~\ref{sec:discussion}) and demonstrates that qualitative information can be extracted from the transient reflectance trace with no prior information on the exciton dipole moment and without a perfect knowledge of the IR pulse. This result motivates the choice for a subsequent ePIE application to polish the reconstruction of both object and probe.

\subsection{\label{subsec:IIstep}Second stage: ePIE}
\begin{figure*}[htbp]
\centering\includegraphics[width=0.9\textwidth]{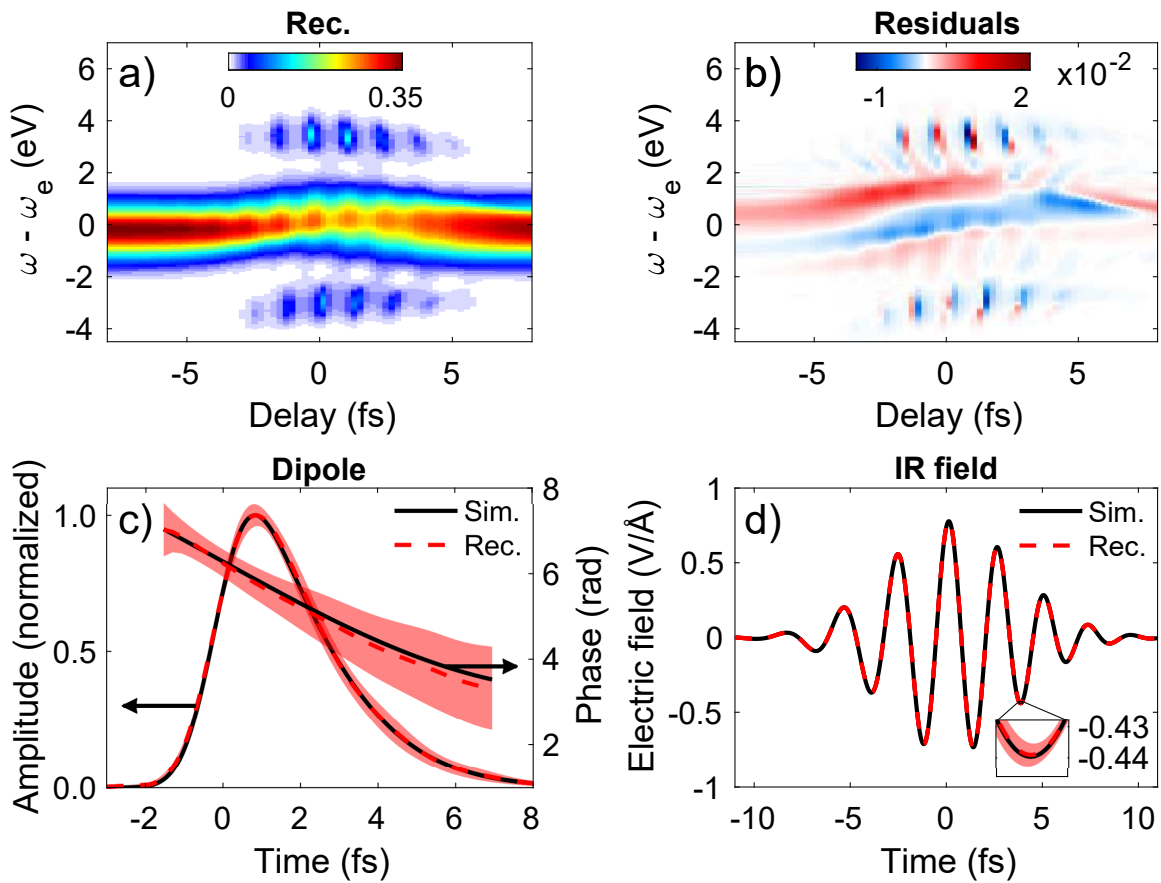}
    \caption{\label{fig:ePIEtraces}Reconstructed reflectivity trace (a) and map of the residuals (b), as a result of the last ePIE stage in the ePIX algorithm. Panels (c) and (d) show the reconstructed dipole moment and IR field, respectively: in both panels the exact solution is presented with solid black curves while the reconstruction results are with dashed red curves. The shaded red area represent the standard deviation calculated over the four different reconstructions performed. The recovery of the IR field resulting from the application of ePIE leads to a more accurate reconstruction of the reflectivity trace, as can be seen by comparing the residuals with the ones in Fig.~\ref{fig:PIE1}(d).}
\end{figure*}


If the IR pulse is not known with enough accuracy, as for the present case, the ePIE stage might be needed to refine the results. The output of the previous step becomes the new initial guess for this stage, where the algorithm updates object and IR field as described in Section~\ref{sec:algorithm}. The results obtained after approximately 1000 iterations of the ePIE step are displayed in Fig.~\ref{fig:ePIEtraces}. The agreement between the simulated and the reconstructed reflectivity traces (Fig.~\ref{fig:ePIEtraces}(a)) is improved by one order of magnitude (compare the residuals in Fig.~\ref{fig:PIE1}(d) with the ones of Fig.~\ref{fig:ePIEtraces}(b)). Moreover, the better accuracy in the reconstruction of the reflectivity trace is translated to a more precise retrieval of both the excitonic dipole and IR electric field time evolution (Figs.~\ref{fig:ePIEtraces}(c) and \ref{fig:ePIEtraces}(d), respectively). Table~\ref{tab:field} summarizes the values of the IR field parameters retrieved by the last ePIE stage employed in ePIX. The good agreement between reconstructed and simulated parameters confirms the ability of the ePIE stage in recovering the correct IR field.
\begin{table}[htbp]
    \caption{\label{tab:field}%
    Simulated and retrieved values of the IR field parameters after the last ePIE stage in the ePIX algorithm. The indicated errors are the standard deviations, arising from the results of four different reconstructions.
    }
    \begin{ruledtabular}
    \begin{tabular}{lcc}
    \textrm{Parameter}&
    \textrm{Simulated}&
    \textrm{ePIE retrieval}\\
    \colrule
    FWHM (fs)&
    8&
    7.99\,$\pm$\,0.046\\
    $\Phi$ (rad)&
    1.26&
    1.25\,$\pm$\,0.013\\
    CR (fs$^{-2}$)&
    0.05&
    0.051\,$\pm$\,0.0035\\
    \end{tabular}
    \end{ruledtabular}
\end{table}

\subsection{\label{subsec:IIIstep}Final PIE polishing}

\begin{figure*}[htbp]
\centering\includegraphics[width=0.9\textwidth]{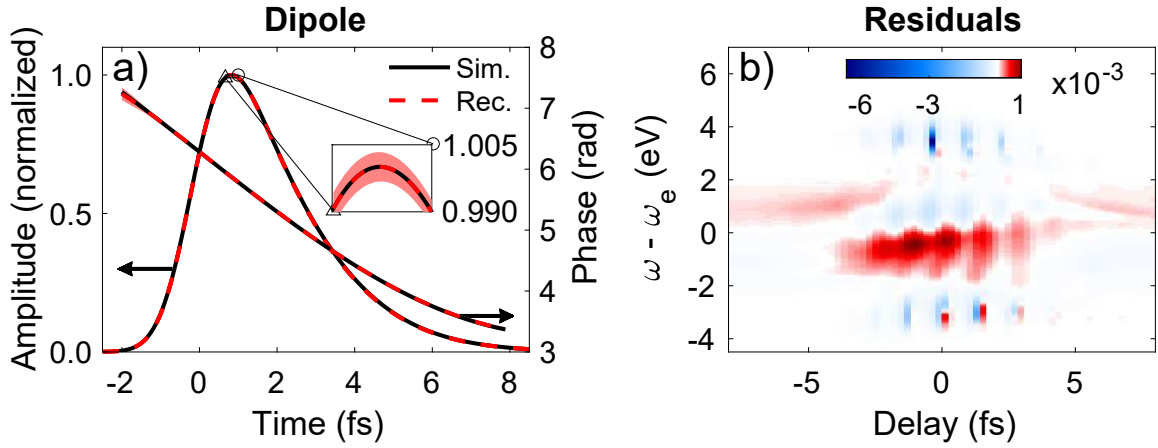}
    \caption{\label{fig:finaldip} (a) Exact (solid black) and final ePIX reconstructed (dashed red) dipole moment $O(t)$ in amplitude and phase. (b) Associated map of the residuals. Both amplitude and phase are correctly recovered with very small uncertainty. As the last stage in the ePIX algorithm does not update the IR field, the final IR reconstruction is the same as Fig.~\ref{fig:ePIEtraces}(d).
}
\end{figure*}
Despite the outstanding results obtained after the ePIE stage discussed above, we found beneficial to close the ePIX algorithm with the application of a PIE stage in order to further refine the excitonic dipole in view of the more accurate IR guess obtained. Following the ePIX iterative approach, this final step takes as input parameters the output of the previous step. The IR electric field is fixed while the excitonic dipole and polarizability are updated. After about 1000 iterations of the PIE code, the maximum discrepancy of the reconstructed dipole is 0.42\% of the peak amplitude and 14 mrad for the phase (see Fig.~\ref{fig:finaldip}(a)), showing the high accuracy of our algorithm.

\section{\label{sec:discussion}Discussion}
In this Section, we compare the exciton physical parameters obtained at each ePIX stage. To this end, for each of the three steps described in Section~\ref{sec:results}, we fit the resulting exciton dipole with the 1D fitting curve given by Eq.~(\ref{eq:dipole}). The retrieved parameters are summarized in Tab.~\ref{tab:allresults}, where the reported errors are the standard deviations arising from the results of four different reconstructions. 
\begin{figure}[htbp]
\centering\includegraphics[width=0.5\textwidth]{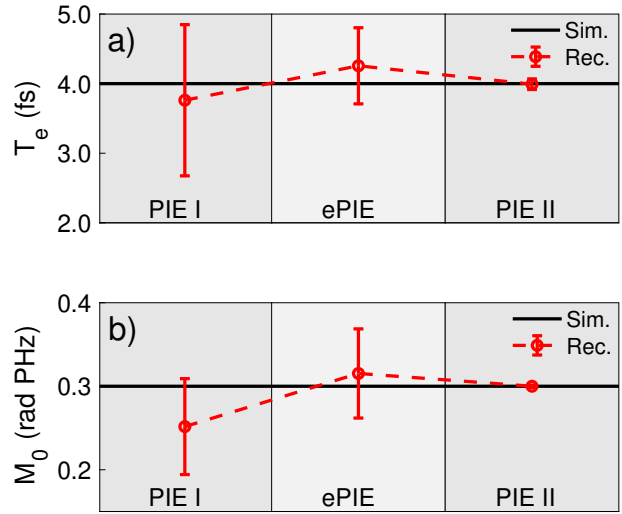}
    \caption{\label{fig:parameters}Reconstructed Auger lifetime (a) and phonon coupling (b) as a function of the reconstruction stage in ePIX. Note that the average value of each parameter approaches the simulated value, and the error bar narrows. A similar behaviour is observed for all other parameters.
}
\end{figure}
Notice that the right order of magnitude for the parameters is retrieved already at the first reconstruction step (PIE I, third column in Tab.~\ref{tab:allresults}), even if the IR field is not accurately known. Indeed, the estimated values agree with the exact theoretical quantities within less than 20\% relative error (except for $\gamma$, which clearly requires further reconstruction steps to be more accurately recovered). The subsequent ePIE step corrects the IR estimation, considerably refining the exciton parameters (fourth column in Tab.~\ref{tab:allresults}) which now agree with the exact values within a 10\% relative error. The last step (PIE II, fifth column in Tab.~\ref{tab:allresults}) reduces the reconstruction uncertainty, improving the overall accuracy. We found the final output parameters to agree with the theoretical values with high degree of accuracy. To better display the effect of the subsequent reconstruction steps, in Fig.~\ref{fig:parameters} we show the evolution of the Auger lifetime (Fig.~\ref{fig:parameters}(a)) and phonon coupling (Fig.~\ref{fig:parameters}(b)) through ePIX. While the second ePIE step changes significantly the parameter estimation, the third step does not correct the mean value significantly, but rather increases the reconstruction accuracy as marked by the smaller error bars.
These results not only prove the transient reflectance trace to be sensitive to the exciton properties, but also that the ptychographic-based algorithm we developed is well-founded, demonstrating to be a valuable tool for the investigation of exciton dynamics in transient reflectance measurements. To further check this, we applied the ePIX method to various transient reflectance traces generated with different physical parameters. We found our approach to converge in all cases with the same accuracy discussed here. If some of the physical parameters are known \textit{a priori}, both ePIX convergence and accuracy are boosted. What we discussed in this work is a ``worst-case scenario'', where no physical parameter is known and an accurate IR time characterization is not available.

For the sake of simplicity, all the calculations and reconstructions reported here include only the exciton ground state contribution to the transient reflectivity trace. Nevertheless, in a real experiment, the transient reflectivity trace $R(\omega, \tau)$ is more likely to contain features related to several excitonic transitions, including dark states which become optically active under the effect of the driving IR field. While this hinders the direct application of the model described here to experimental traces, the ePIX approach remains valid as it does not assume a particular functional shape for the object $O(t)$ (excitonic dipole) and can therefore be extended to account for different excitonic transitions. Of course, in such a case, the model used to perform the 1D fit and extract the physical parameters of interest needs to be changed accordingly. It is worth noticing that, in case of congested spectra, the ePIX accuracy might be reduced because of the large number of free parameters to be reconstructed. Nevertheless, we expect our approach to perform better than a 2D fit of the transient reflectance trace. Indeed, with the ePIX method we can unambiguously determine all the parameters that govern the ultrafast dynamics of a core exciton and its interaction with the crystal, such as its Auger lifetime or the coupling with the phonons. The understanding of such properties is crucial for shedding new light on the behaviour of solid-state materials and assessing their features with unprecedented accuracy. Therefore, our reconstruction algorithm, together with the experimental methods of the attosecond science, may serve the purpose of widening the knowledge about light-matter interaction, paving the way for a full comprehension and control of its optoelectronic properties, up to the petahertz regime.

\section{\label{sec:conclusions}Conclusions}

In this work, we presented a novel approach, named ePIX, to recover the ultrafast core-exciton dynamics from transient reflectivity measurement. This method is based on a reformulation of the PIE and ePIE algorithms in an iterative approach. By applying ePIX to a simulated transient reflectivity trace, computed in conditions similar to what reported in Ref.~\cite{lucchini2021unravelling}, we prove that the method is capable to accurately reconstruct the exciton dynamics even with no \textit{a priori} knowledge of any of the physical parameters which describe the system and with little knowledge of the IR driving electric field. The results are compared with a more common 2D fitting procedure, showing superior degree of accuracy and sensitivity against the complexity of the physical phenomena under examination. Finally, as the approach we proposed is based on an atomic-like description, ePIX can be applied beyond the field of solid-state physics, becoming a valuable tool for studying the ultrafast dynamics in different physical scenarios.

\begin{acknowledgments}
This project has received funding from the European Research Council (ERC) under the European Union’s Horizon 2020 research and innovation programme (grant agreement No.~848411 title AuDACE). ML and GI further acknowledge funding from MIUR PRIN aSTAR, Grant No.~2017RKWTMY. MN acknowledges funding from MIUR PRIN, Grant No.~20173B72NB. We acknowledge support from Laserlab-Europe EU-H2020 GA No. 871124.
\end{acknowledgments}

\bibliography{apssamp}

\end{document}